# The drag force distribution within regular arrays of cubes and its relation to cross ventilation – theoretical and experimental analyses

Riccardo Buccolieri[a, b*], Mats Sandberg[b], Hans Wigö[b], Silvana Di Sabatino[c]

[a]Dipartimento di Scienze e Tecnologie Biologiche ed Ambientali, University of Salento, S.P. 6 Lecce-Monteroni, 73100 Lecce, Italy

[b]Faculty of Engineering and Sustainable Development, University of Gävle, SE-80176 Gävle, Sweden

[c]Department of Physics and Astronomy, ALMA MATER STUDIORUM - University of Bologna, Viale Berti Pichat 6/2, 40127 Bologna, Italy

*Corresponding author*: Riccardo Buccolieri
Email: riccardo.buccolieri@unisalento.it
Tel. +39 0832 297 062; fax +39 0832 297 061

**Abstract**

A novel set of wind tunnel measurements of the drag force and its spatial distribution along aligned arrays of cubes of height $H$ and planar area index $\lambda_p$ (air gap between cubes) equal to 0.028 ($5H$) to 0.69 ($0.2H$) is presented and analysed. Two different types of measurements are compared: one type where the drag force is obtained using the standard load cell method, another type where the drag force is estimated by measuring the pressure difference between windward and the leeward façades. Results show that the drag force is nearly uniformly distributed for lower $\lambda_p$ (0.028 and 0.0625), it decreases up to 50% at the second row for $\lambda_p$=0.11, and it sharply decreases for larger $\lambda_p$ (from 0.25 to 0.69) where the force mostly acts on the first row. It follows that for the lowest $\lambda_p$ the drag force typically formulated as a drag area corresponds to the total frontal area of the array, whereas for large $\lambda_p$ the drag area corresponds to the area of the first row. By assessing the driving pressure for ventilation from the drag force, the analysis is extended to estimate the cross ventilation as an example of application of this type of measurements.

**1. Introduction**

Several studies of the modification of the flow due to an array of obstacles exposed to a boundary layer flow have been carried out in the past. These studies can be broadly subdivided into two main categories. One category is based on the classical fluid mechanical approach where the focus is on the evolution of the approaching velocity profile when passing through the array. Usually the array is long enough that an equilibrium is established. An excellent recent example of this approach is given by Thomas et al. (2017), where references to similar studies are provided.

The other category is where the main concern is the ventilation of the array taking into account individual buildings. In this case the focus is on understanding how both the whole array interacts with the approaching wind and how the individual buildings interact with the airflow through the array. The overarching goal of this paper is to contribute to this understanding. This paper is essentially a continuation of our previous work (Buccolieri et al., 2017), where the total drag force was measured in a wind tunnel by using a standard load cell. The focus was to study the effect on the drag force of different building packing density of an array consisting of buildings (represented by cubes) of equal size and shape. A novel method for assessing the distribution of the drag force was introduced by formulating the total drag force as a drag area and then matching this area with the physical façade area of the buildings. One scope here is to validate this method by directly measuring the distribution of the drag force within the array using the same standard load cell method. To provide insight into the quality of the measurements and confidence for the obtained results, the drag force is also assessed by an independent method based on measurements of the surface pressure at the windward and leeward façades of the buildings.

The results are analysed from the perspective that the introduction of an array in a given turbulent flow is a perturbation of a reference case which we have chosen to be the isolated cube. The introduction of new buildings to the reference case increases the resistance (drag) and less air (less flow rate) penetrates into the array compared to the reference case. Therefore it takes a longer time for the approaching air to pass through the array. This delay has been quantified by Antoniou et al. (2017) for a region within Nicosia in Cyprus by predicting the mean age of air in the city. The delay has a consequence for air quality because the local mean age of air is directly proportional to the concentration in case of homogeneous emissions, see Eq.10 in Buccolieri et al. (2010). Low flow rate penetrating into the array means that less flow available for ventilation of the buildings.

The argument above helps clarifying that the drag is an important parameter linked to both air quality within the array and ventilation potential available for buildings in a given neighbourhood or city. There are several wind tunnel studies reporting on pressure distribution measurements at building façades as well as estimate of the drag force either pressure-derived or using a balance. Cheng and Castro (2002) and Cheng et al. (2007) estimated the drag force on individual cubes within an array by calculating the integral of the pressure difference between the front and the back façades of the cube. Other studies, relevant to the field of wind load on structures, utilized pressure measurements to evaluate the pressure distribution on individual buildings (Kim et al., 2012; Tecle et al., 2013). Zaki et al. (2011) employed surface pressure measurements of the form drag on building arrays featured by both vertical and horizontal randomness as well as different packing densities demonstrating a significant effect of building height variation on aerodynamic parameters when the planar area index is larger than 17%. Li et al. (2015) confirmed the dependence of building shape and position within the array on the drag coefficient by surface pressure measurements. A comprehensive dataset of wind pressure for isolated low- and high-rise buildings, as well as for non-isolated low-rise buildings, is available from the Tokyo Polytechnic University (wind.arch.t-kougei.ac.jp/system/eng/contents/code/tpu).

Cheng et al. (2007), Hagishima et al. (2009) and Zaki et al. (2011) directly measured the drag force using a balance. Cheng et al. (2007) showed that the drag force exerted on cube arrays and derived from measured Reynolds stresses can be underestimated by as much as 25% compared to drag directly measured using a balance. Hagishima et al. (2009) measured the drag force directly by a designed floating raft in a wind tunnel to investigate the aerodynamic effects of various building arrays showing that both wind direction and the height non-uniformity of buildings affect aerodynamic parameters significantly.

More recently, few studies have reported measurements of the drag force distribution within building arrays, as done in this paper. Chen et al. (2017) used a standard load cell within arrays consisting of buildings with both the same height and different heights. From the recorded drag force the vertical transport by both advection and turbulence expressed as an exchange velocity was predicted based on methods presented in Bentham and Britter (2003) and in Hamlyn and Britter (2005). Li et al. (2018) reported wind tunnel measurements of the drag distribution within irregular arrays consisting of buildings of different shapes but with the same height. The drag force was estimated from the pressure difference between the windward and leeward façades of the buildings and directly by using floating rafts. They presented drag force distributions and an extensive discussion of estimation of measurements errors and how to conduct this type of measurements is presented.

In this context, the novelties of the present paper are briefly summarised:
- a comprehensive dataset of recorded drag force including both the total drag force of the whole array (shown in Buccolieri et al., 2017) and the distribution of the drag force within the array for a large span of building packing densities has been created. The dataset includes also data obtained from pressure measurements at the wind- and leeward façades of the buildings;
- original analyses are presented showing that an interference between buildings exists and this needs to be quantified in the derivation of the drag force by introducing the wall to wall distance as interference parameter in addition to building area density;
- a further validation of the novel method presented in Buccolieri et al. (2017) for assessing the drag force distribution starting from the drag force measurements of the whole array is provided;
- an application of the potential usefulness of the dataset is provided by estimating the potential for wind-driven cross ventilation. It is shown that measuring the drag force is much simpler than measuring the pressure in several points.

## 2. Description of wind tunnel experiments
### *2.1. The physical models*

Measurements were carried out in a closed-circuit boundary layer wind tunnel with a working section 11m long, 3m wide and 1.5m high at the University of Gävle (Sweden) (Fig. 1a). An isolated cube and seven aligned arrays of cubes of planar area index (ratio between the planar area of buildings and the lot area) $\lambda_p$ from 0.028 to 0.69 were considered. The cube height $H$ was equal to 0.06m. The lot area was a square with a side length of 13$H$ (0.78m) (see Fig. 3a later in the text).

Drag force and pressure measurements were performed separately on one individual cube ("target cube" hereinafter) placed along the middle column of the array (Fig. 1b). Please note than when the number columns was even (as for example for $\lambda_p$=0.0625 in the figure), the measurements were performed along one of the two columns constituting the middle of the array. The target cube was kept fixed on the wind tunnel floor while its position within the array was changed by moving the other cubes. Initially the target cube was located in the first row. In the next step the cubes from the last row were moved to the front so that the target cube was located in the second row of the array. This procedure was repeated until the target cube was positioned in the last row of the array.

The employed geometries are the same as used in our previous paper (Buccolieri et al., 2017), i.e. the lot area was kept constant (the denominator in $\lambda_p$) and the number of cubes was increased to represent neighbourhoods of different $\lambda_p$. In this approach there may exist conditions where flow adjustment occurs and conditions where the flow is still evolving. These experiments intend to reproduce conditions in which the surrounding terrain is almost uniform and there is a considerable transition from a given roughness to a new roughness where the flow within the array may be still evolving.

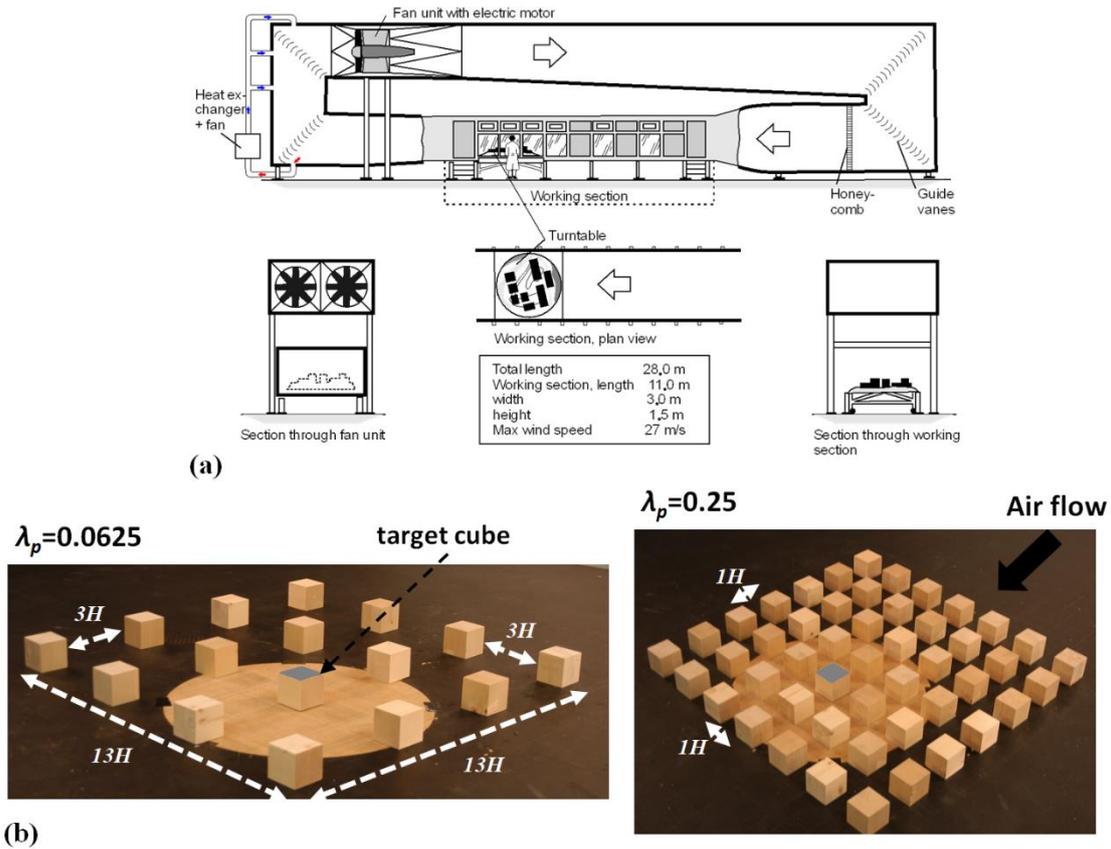

**Fig. 1** a) Sketch of the wind tunnel. b) Example arrays with planar area index $\lambda_p$=0.0625 and 0.25, with indication of air gap between cubes and the target cubes (roof s highlighted in grey) placed in the centre of the turntable along the centre of the arrays.

## *2.2. The boundary-layer flow*

A boundary-layer (BL) flow in the wind tunnel was achieved considering two different conditions for the fetch: (i) the entire fetch was covered with cubes of 0.04m side length representing roughness elements ("BL roughness" hereinafter); (ii) the fetch was smooth with no roughness elements ("BL no roughness" hereinafter). The distance between the final row of roughness elements and the front of the lot area was approximately 0.4m. The roughness area in the working section of the wind tunnel had a total length of 8m made of spires in the first part and then of 0.04m cubes (roughness elements).

The boundary layer thickness $\delta$ was about 0.15m (2.5*H*) in the "BL no roughness" case and about 0.8m (13.3*H*) in the "BL roughness" case, as estimated by taking the height at which the velocity was equal to 99% of the free stream velocity. The blockage coefficient ($\Phi = A_{\mathrm{model,proj.}}/A_{wind\_tunnel}$, where $A_{model,proj.}$ [m$^2$] is the projected area of the cube along the main wind direction and $A_{wind\_tunnel}$ is the cross-sectional area of the measurement section in the wind tunnel), was about 0.1% which fulfils the requirements of the VDI 3783 guidelines (2004).

The experiments were performed with one reference wind velocity $U(H)$ [ms$^{-1}$] at the cube height $H$, corresponding to 500 revolutions per minute (rpm) of the fan that drove the flow in the wind tunnel. The independence of the drag force on the reference velocity was tested in the previous work (Buccolieri et al., 2017). The error is within ±5% of the measured value. Undisturbed mean velocity and relative turbulence intensity profiles approaching the array (which are in equilibrium with the roughness in the fetch), both normalised by the corresponding value at $H$=0.06m up to $z/H$=2.5, are (Fig. 2):

$$\frac{U(z)}{U(H)} \approx \left(\frac{z}{H}\right)^{0.16} \qquad (1)$$

$$\frac{I(z)}{I(H)} \approx \left(\frac{z}{H}\right)^{-0.06} \quad \text{(BL roughness)} \qquad (2)$$

$$\frac{I(z)}{I(H)} \approx \left(\frac{z}{H}\right)^{-0.46} \quad \text{(BL no roughness)} \qquad (3)$$

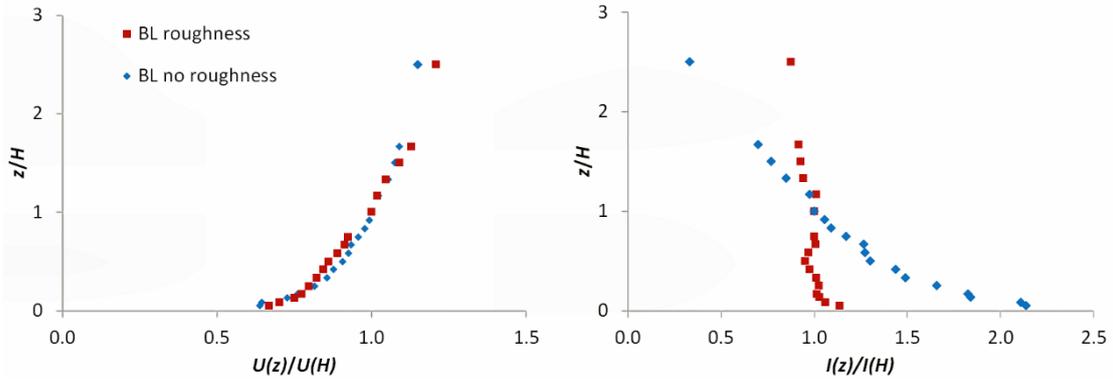

**Fig. 2** a) Boundary layer (BL) wind velocity (left) and relative turbulence intensity (right) incoming profiles in the wind tunnel for rpm = 500. The profiles have been fitted using power laws in Eqs. 1, 2 and 3.

The velocity was measured with a TSI hot-film anemometer in the middle of the empty circular disk where the target cube was attached (see subsection 2.3). The turbulence intensity was calculated as the standard deviation of the velocity fluctuations divided by the mean velocity. Tab. 1 summarizes the cases investigated.

**Table 1**
Summary of all test cases investigated in the wind tunnel. *U(H), I(H)* and *Re* are the incoming mean flow velocity and turbulence intensity at building (cube) height *H* and the Reynolds number, respectively.

| No. of cubes | $\lambda_p$ | Air gap between buildings (in the transversal and longitudinal directions) | Centre to centre distance between buildings (in the transversal and longitudinal directions) | $U(H)$ [ms$^{-1}$] - $I(H)$ [%] – $Re$ [-] | |
|---|---|---|---|---|---|
| | | | | **BL roughness** | **BL no roughness** |
| 1 | - | Isolated cube | | 5.2 - 27.6 - 20,800 | 9.5 - 7.1 - 38,000 |
| 3x3 | 0.028 | 5*H* | 6*H* | | |
| 4x4 | 0.0625 | 3*H* | 4*H* | | |
| 5x5 | 0.11 | 2*H* | 3*H* | | |
| 7x7 | 0.25 | 1*H* | 2*H* | | |
| 9x9 | 0.44 | 0.5*H* | 1.5*H* | | |
| 10x10 | 0.56 | 0.33*H* | 1.33*H* | | |
| 11x11 | 0.69 | 0.2*H* | 1.2*H* | | |

*Note: the Reynolds number, based on the height of the cube H and the incoming undisturbed reference flow velocity U(H) at cube height H, is lower in the "BL roughness" case as a consequence of the smaller U(H).*

*2.3. Drag force measurements*

The drag force $F_D$ (balance) acting on the individual target cube was directly measured using the standard load cell method described in Buccolieri et al. (2017). Specifically, the target cube was connected to the load cell via two thin rods that went through a small opening in the turn table. There was an air gap of 1mm between the cube and the turn table. The load cell was mounted on a stable tripod standing on the floor of the laboratory hall (Fig. 3) so that the cube was mechanically isolated from the wind tunnel and the measured force was only due to air resistance. To measure the drag force distribution within the array, the force was singularly measured on the cubes located along the centre of the array (along the wind direction).

In the standard load cell the horizontal force, caused by the air movement, was transformed into vertical tensile and compressive force at its edges. Here it was Vetek 108AA with glued strain gauges (Vetek, 2016) which measured the forces and provided an electrical output signal. The signal was then amplified through the Amplifier, converted to digital through the 16 bit AD-converter and finally read by the Lab View program (Fig. 3). In the program the signal offset (zero) and gain could be adjusted before further processing. The signal from the load cell was sampled at 1000Hz and then a mean value was calculated every second. Due to turbulence the measuring signal still fluctuated and further signal processing was necessary. To obtain stable measurements a sliding average was considered using 60s. The force was read when the sliding average was stable.

The load cell measured the force along the flow direction since it was mounted in parallel with the main wind flow. The load cell had an internal compensation that balances out the torque. Therefore, it measured the net force in the flow direction regardless of where the force acted on the cube. The accuracy was tested and the total measurement uncertainty is specified as the reading ± 7%. Further details, including the calibration procedure, are given in Buccolieri et al. (2017).

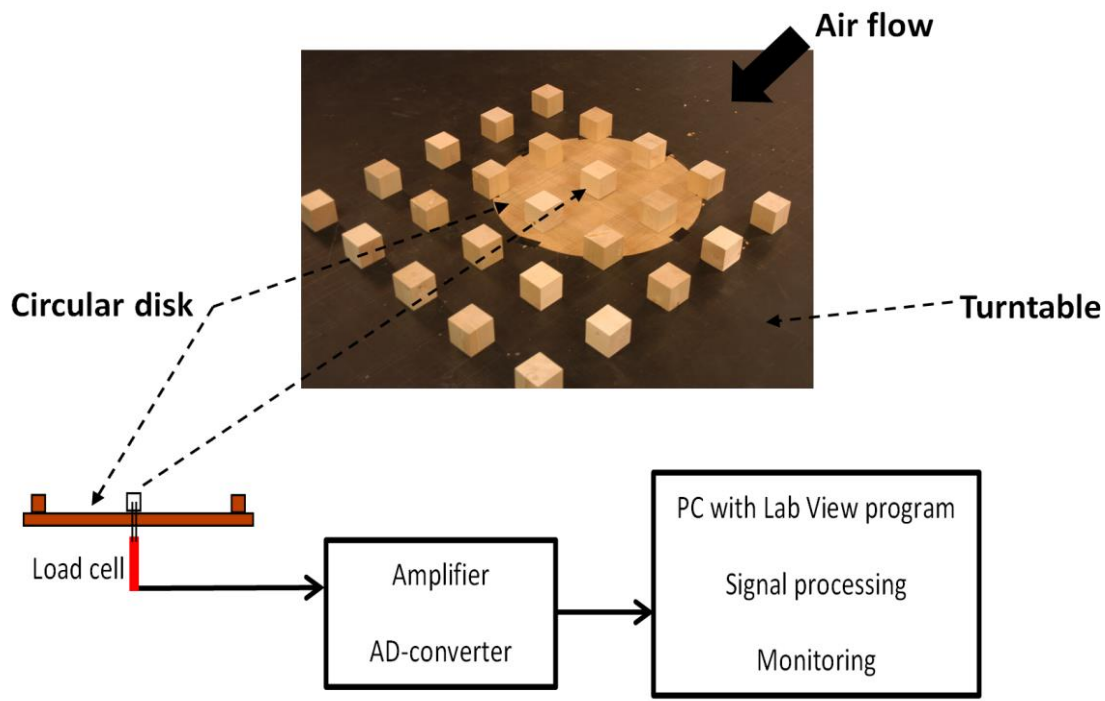

**Fig. 3** Pictures showing an example of the array built in the wind tunnel and attached to the circular disk, with indication of the target cube connected to the load cell.

*2.4. Pressure measurements and drag force estimation*

The static pressure at windward and leeward sides was measured via pressure taps placed at the façades of the target cube (Fig. 4a,b). The diameter of the tap openings was 0.8mm and the opening was oriented perpendicular to the wall. The pressure was measured only at one half of the façades. All pressure taps were connected to a multiplexer (scanner valve), which transferred each pressure to the Furness FCO12 pressure transducer. The signal was sampled with 1000Hz and the final reported pressure was the average over 30s. The area was then divided into 40 sub-areas ($A_i$ with $i$=1 to 40) according to where the taps were located (Fig. 4c).

The drag force, acting perpendicular on the windward side of the target cube, was then calculated as follows:

$$F_{D\_windward}(\text{pressure}) = \sum_{i=1}^{n} p_i \times A_i \tag{4}$$

where the measured pressure $p_i$ was assumed to be constant over the entire sub-area $A_i$. On the leeward side of the target cube (area $A_{leeward}$) the pressure distribution was almost uniform so the force was calculated from the average pressure $p_{average}$ as follows:

$$F_{D\_leeward}(\text{pressure}) = p_{average} \times A_{leeward} \tag{5}$$

Sensitivity tests using Eq. 4 for calculating $F_{D\_leeward}(\text{pressure})$ showed a percentage difference lower than 2%.

The total drag force $F_D(\text{pressure})$ acting on the target cube, along the flow direction, was finally calculated as:

$$F_D(\text{pressure}) = F_{D_{windward}}(\text{pressure}) - F_{D_{leeward}}(\text{pressure}) \tag{6}$$

The direction of the force was determined by the sign of the pressure.

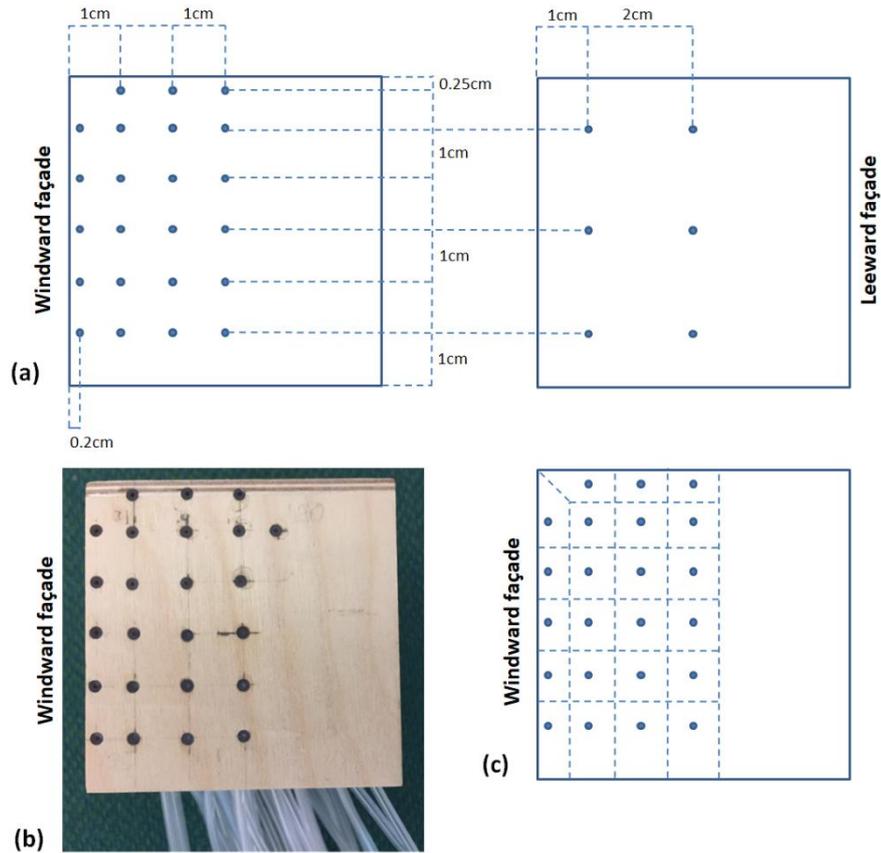

**Fig. 4.** a, b) Pressure taps position at windward and leeward façades of the target cube. c) Sub-areas of the windward façade employed for the calculation of the drag force via the pressure-derived method.

It should be noted that Eq. 6 represents an integration based on a finite number of sampling points. One can expect the result to be dependent on the number of sampling points and the location of the sampling points, since due to the acceleration of the air flow towards the edges, on the windward side of the building the static pressure varies a lot across the surfaces (see. Fig. 9 later in the text). Therefore, as shown in Fig. 4a, after several tests (not shown here) with different number of sampling points on both the windward and leeward façades, some pressure taps were also placed near the edges of the wall, resulting in good agreement between the calculated (by the pressure-derived method) and measured (with the balance) drag force (see Fig. 7 later in the text).

## 3. Drag area and cross ventilation
### *3.1. The drag area*

The drag coefficient based on the drag force $F_D$ generated by the target cube, the reference velocity $U(H)$ and the physical frontal area of the cube $A_{Cube}$ is defined as:

$$C_D = \frac{F_D}{\frac{1}{2}\rho U(H)^2 A_{Cube}} \quad (7)$$

With the aim of determining the effective area of cubes generating the drag force, we use the definition of the drag area (Buccolieri et al., 2017) by setting the drag coefficient equal to 1 and solving for the area. The drag area of the target cube ($A_D$) becomes:

$$A_D = \frac{F_D}{\frac{1}{2}\rho U(H)^2} \quad (8)$$

The rationale for introducing a drag area is that it can be compared with the physical area of the cubes in the array. The drag coefficient becomes the ratio between the drag area and the physical area of the cube:

$$C_D = \frac{A_D}{A_{cube}} \quad (9)$$

Regarding the magnitude of the drag area there are two extreme cases:

*Long distance between the cubes (low $\lambda_p$) – No interference between the cubes*

The cubes behave as independent bodies and the total drag force acting on the whole array ($F_{D\_total}$) is equal to the number of cubes $N$ multiplied by the drag force $F_D$ generated by the single target cube:

$$F_{D\_total}(\text{for low } \lambda_p) = N F_D \quad (10)$$

which, by Eq. 8, leads to the total drag area of the array ($A_{D\_total}$):

$$A_{D\_total}(\text{for low } \lambda_p) = \frac{F_{D_{total}}}{\frac{1}{2}\rho U(H)^2} = N \frac{F_D}{\frac{1}{2}\rho U(H)^2} = N A_D \quad (11)$$

*Short distance between the cubes (large $\lambda_p$) – Strong interference between the cubes)*

The extreme case is when total force $F_{D\_total}$ is on the cube at the front ($F_{D\_front\_cube}$) which, by Eq. 8, leads to total drag area:

$$A_{D\_total}(\text{for large } \lambda_p) = \frac{F_D}{\frac{1}{2}\rho U(H)^2} = \frac{F_{D\_front\_cube}}{\frac{1}{2}\rho U(H)^2} = A_{D\_front\_cube} \quad (12)$$

To generalize, if $F_D^i$ is the drag force generated by cube $i$ within the array and $A_D^i$ is the corresponding drag area, the total drag force exerted by the whole array is equal to:

$$F_{D\_total} = \sum_{i=1}^{N} F_D^i \quad (13)$$

and the total drag area $A_{D\_total}$ (see Eq. 8) becomes:

$$A_{D\_total} = \frac{\sum_{i=1}^{N} F_D^i}{\frac{1}{2}\rho U(H)^2} = \sum_{i=1}^{N} A_D^i \tag{14}$$

By dividing the total drag area by the total physical frontal area for each $\lambda_p$, $\frac{A_{D\_total}}{\text{Frontal area}}$, it can be evaluated which is the most appropriate reference area (drag area) to be used for the calculation of the drag coefficient $C_D$, i.e. when the ratio tends to one.

### *3.2 Assessment of wind-driven cross ventilation based on the drag force*

Wind-driven natural ventilation of buildings occurs either as cross ventilation or single-sided ventilation (Etheridge and Sandberg, 1996). Single- sided ventilation occurs when all openings are located on one side (Warren, 1977). Cross ventilation occurs when there are openings on different sides of a building.

Cross ventilation has been studied by several authors using different approaches. Some examples are provided by Karava et al. (2007, 2011), Chu and Chiang (2014) and Shetabvish (2015). More recently Shirzadi et al (2018) have investigated cross ventilation for buildings embedded in arrays using computational fluid dynamics. When the pressure inside the building is uniform (Kobayashi et al., 2010), cross ventilation is assumed to be driven by the difference of the static pressure between outside and inside of the building which generates a velocity through the openings. If the building is a bluff body, the drag force is generated by the pressure difference between the windward and the leeward façades .

In the present paper, in order to assess the potential for cross ventilation it is assumed here that the cubes in the investigated arrays are provided with two openings opposite to each other (Fig. 5) and for simplicity the area of both openings are taken to be equal. By entrainment into the air stream flowing into the room there is a gradual expansion of the air stream. If the distance *L*, to the leeward wall, is larger than about six times the linear dimension of the opening, the cross section of the air flow have expanded so it is larger than the opening in the leeward wall. Then the whole air stream cannot continue straight through the opening on the opposite side. This is the prerequisite for the pressure inside the building to be uniform.

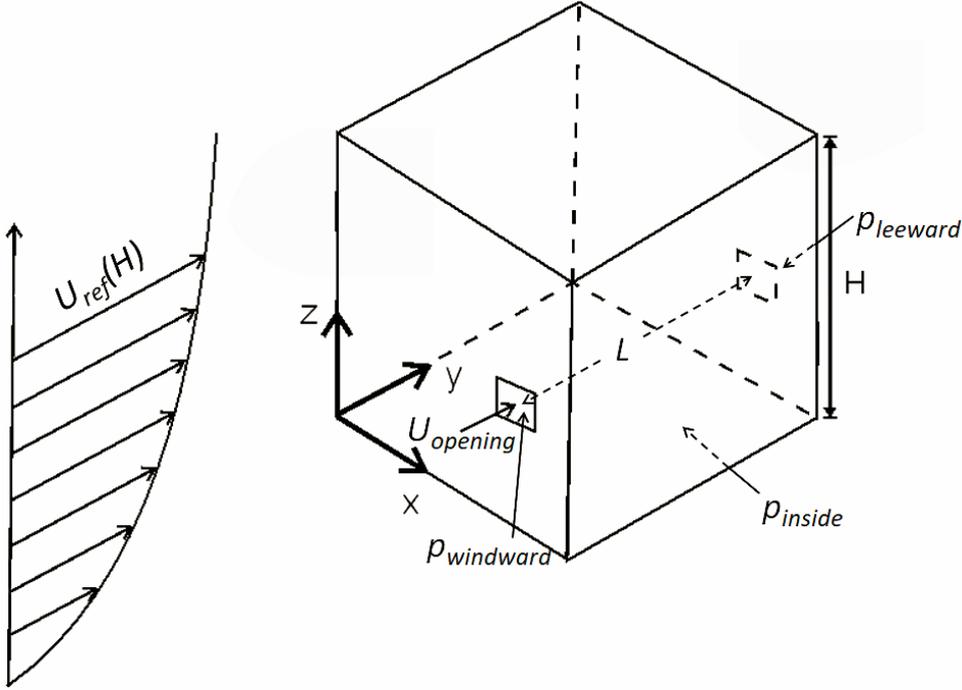

**Fig. 5.** A cube with two openings located opposite to each other.

The flow rate $Q_{opening}$ [m³s⁻¹] through the opening is, by assuming that the velocity profile is uniform across the opening, given by:

$$Q_{opening} = A_{opening} U_{opening} \qquad (15)$$

The velocity in the opening, $U_{opening}$, is driven by the pressure difference $|\Delta \bar{p}_{Out-In}|$ between outside of the building and the inside of the building according to the orifice equation (Etheridge and Sandberg, 1996):

$$U_{opening} = C_{discharge} \sqrt{\frac{2|\Delta \bar{p}_{Out-In}|}{\rho}} \qquad (16)$$

where $C_{discharge}$ is the discharge coefficient which takes into account several factors as e.g. area contraction of the stream tube when passing through the opening and losses.

For the opening on the windward façade $\Delta \bar{p}_{Out-In} = (\bar{p}_{windward} - \bar{p}_{inside})$ and for the opening on the leeward façade $\Delta \bar{p}_{Out-In} = (\bar{p}_{inside} - \bar{p}_{leeward})$. The unknown quantity is the pressure $\bar{p}_{inside}$ within the cube which we assume to be constant within the cube. This condition is fulfilled if the openings are not too large. The pressure variation inside a building related to the size of the openings is shown in Kobayashi et al (2010). This pressure inside the cube is obtained from the flow balance that dictates that the flow rates through both openings are the same. For simplicity, we set the discharge coefficients for the openings on the windward and the leeward façades the same. This implies that the pressure difference across both openings is the same and the pressure within the cube is equal to the average of the pressure on the windward and the leeward façades. Therefore, the pressure difference across both openings becomes equal to half the pressure difference between the leeward and the windward side:

$$\Delta \bar{p}_{Out-In} = \frac{1}{2}(\bar{p}_{windward} - \bar{p}_{leeward}) \tag{17}$$

This pressure difference can be assessed from the drag force $F_D$(balance) measured with the standard load cell and the physical surface area of the cube $F_D$(balance) as follows:

$$(\bar{p}_{windward} - \bar{p}_{leeward}) \approx \frac{F_D(\text{balance})}{A} \tag{18}$$

Inserting this into Eq. 16 the velocity through the opening $U_{opening}$ in relation to the reference velocity $U(H)$ can be written as:

$$\frac{U_{opening}}{U(H)} = \frac{C_{discharge}\sqrt{\frac{F_D(\text{balance})}{\rho A}}}{U(H)} \tag{19}$$

The discharge coefficient $C_{discharge}$ reported in the literature varies a lot. This is due to that apart from different conditions at the tests, the discharge coefficient takes into account many factors. In many practical applications the discharge coefficient is an adjustment factor that links flows at complex conditions to the orifice equation. According to Karava et al. (2004) the discharge coefficient varies between 0.14-0.65 and according to Cruz and Viegas (2016) it varies between 0.47-0.81. In our assessment a precise estimation of the discharge coefficient is not available and thus we set it equal to 1 because the whole analysis is only an order of magnitude estimate. This implies that we, if the orifice equation is valid, overestimate the potential. After setting the discharge coefficient equal to 1, Eq. 19 leads to:

$$\frac{U_{opening}}{U(H)} = \frac{\sqrt{\frac{F_D(\text{balance})}{\rho A}}}{U(H)} = \frac{1}{\sqrt{\rho A}} \frac{\sqrt{F_D(\text{balance})}}{U(H)} \tag{20}$$

Please note that if an estimate of the discharge coefficient is available, this can the easily be taken into account by multiplying Eq. 20 by the new discharge coefficient and the results will be qualitatively similar to those presented in subsection 4.4 (see Fig. 10 later in the text).

Eq. 20 is used here to evaluate the potential for cross ventilation driven by a static pressure difference. The façade area $A = H^2$ is constant while the reference velocity at roof height $U(H)$ is dependent on the type of boundary layer generated (roughness elements in the fetch or not). According to Eq. 20 the velocity in the opening $U_{opening}$ approaches zero when the force approaches zero, which occurs when the difference in mean pressure (Eq. 18) approaches zero. Now other mechanisms for exchange of air between the interior of the building and the ambient come into play (Haghighat et al., 1991). One mechanism is the penetration of turbulent eddies through the openings and another one is flow driven by pressure fluctuations. Pressure fluctuations are generated by e.g. vortex shedding (Zu and Lam, 2018). These mechanisms give rise to penetration phenomena with strong variations in time and therefore the exchange between the indoor and the ambient cannot always be estimated from the flow rate in the opening based on the velocity field (air exchange rate). Instead the exchange must be based on an exchange of a passive contaminant present indoor. This exchange is retained from concentration data. This exchange rate is the purging flow rate (Etheridge and Sandberg, 1996) and is maximized by the air exchange rate. An example of predicting the purging flow rate from concentration data can be seen in Kobayashi et al (2018).

It should be noted that there is a direct relation between the velocity in opening $U_{opening}$ according to Eq. (20) and the assessment of the in-canopy velocity $U_C$ according to Bentham and Britter (2003) relation (2):

$$U_C = \sqrt{2} U_{opening} \tag{21}$$

This relation is obtained from the relation (2) in Bentham and Britter (2003) by substituting the left hand side with the drag force and in the right hand side setting the drag coefficient equal to 1. Eq. 21 implies that it is possible to read off the variation in in-canopy velocity from the variation in the velocity through an opening and vice versa.

## 4. Results
### *4.1. Evaluation of the drag distribution*
### *4.1.1 Interaction between the approaching flow and the array*

Fig. 6a shows a sketch of the flow pattern when the wind approaches the array. For the approaching wind the array constitutes a resistance consisting of a blockage by cubes generating a drag force, while for the air stream passing through the street canyons the resistance is generated by a friction against the surfaces forming the street canyons. At the frontal façades of cubes there are stagnation points and for the air stream passing along the street canyons there are corresponding retardment points defined as the points with the highest static pressure (Sandberg et al., 2004). Due to the increased resistance only a fraction of the approaching flow can penetrate into the array because the street canyons have a lower flow capacity (Hang et al, 2010) than the surrounding non occupied terrain. The fraction (dashed line in the figure) that does not entrain into the array continues above the array. At the downstream end of the array there is a corresponding change from a higher to a lower resistance that causes the flow capacity downstream to increase. This change generates a downward flow at the downstream end of the array. Fig. 6b shows the evolution of the static, dynamic and total pressure in the approach flow continuing through the street canyon.

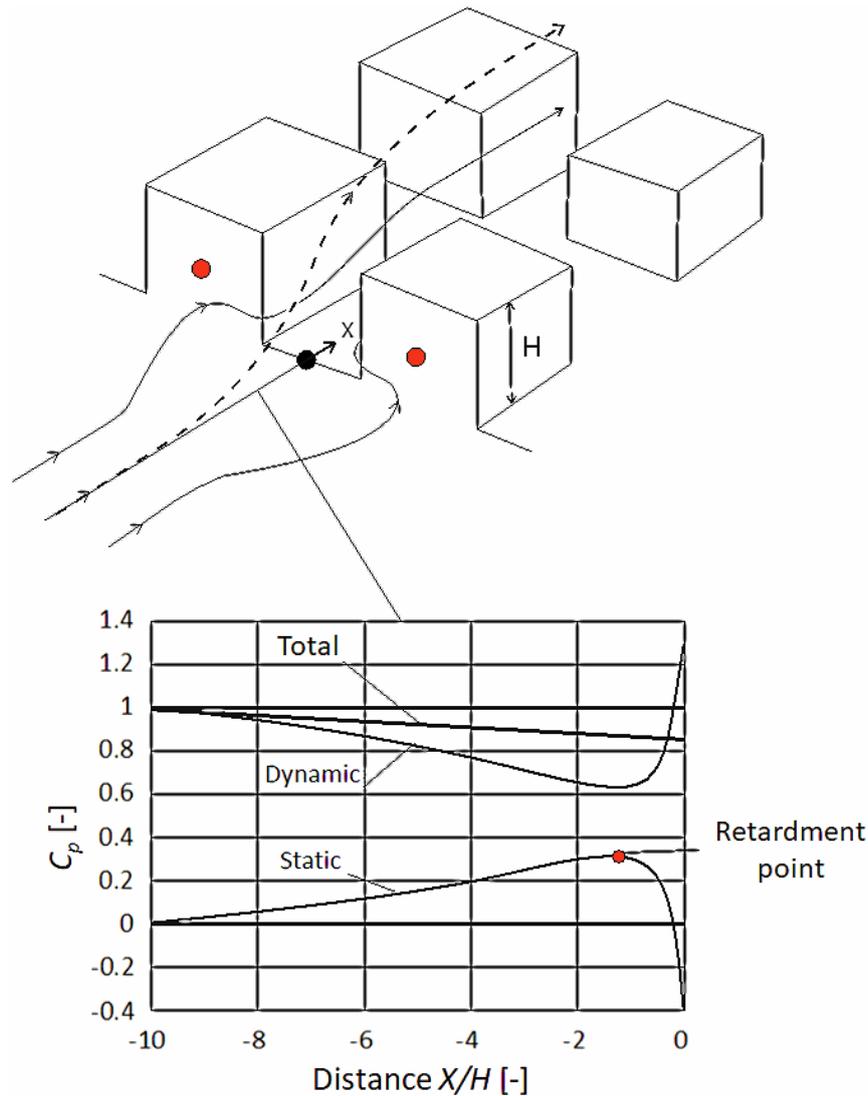

**Fig. 6.** Flow towards the cubes in the front row. a) Sketch of the flow pattern and b) pressure coefficient by static, dynamic and total pressure along the line indicated in a) (adapted from Sandberg et al., 2004).

*4.1.2 Interaction between the cubes and the boundary layer*

For an isolated cube a shear boundary layer is developed and by flow separation a characteristic flow pattern is generated around the cube with a wake on the leeward side. With respect to a reference pressure the isolated cube experiences a positive pressure on the windward side and a negative pressure in the wake on the leeward side. The difference between these pressures generates a drag force on the cube. For an array of cubes, the introduction of surrounding cubes creates an air gap around all cubes located within the interior region of the array. The air gaps constrain the flow and consequently the windward and leeward pressures to be changed relative to an isolated cube. There are many physical phenomena that may affect the drag force on individual cubes which depends on their location within the array. The velocity in the wake is less than the free stream velocity and therefore the cube behind the first cube will be sheltered and subsequently exposed to a lower velocity than the first cube. The result is a lower drag force. If the cubes come very close to each other the cube downstream will be fully submerged within the wake of the first cube which results in a force directed opposite to the wind direction. The second cube may also affect the cube upstream by that the location of the point of separation on the sides of the upstream cube is changed. If the cubes are very close to each other there may be no separation from the first cube.

The introduction of surrounding cubes will thus lead to an interference. Based on field measurements and modelling results, Oke (1988) identified three flow regimes for wind direction perpendicular to the street axis in neutral stratification. For widely spaced buildings (aspect ratio between the building height H and the street width W < 0.3), the flow fields associated with the buildings do not interact (isolated roughness flow regime). At closer spacing (0.3<H/W<0.7) the wake behind the upwind building is disturbed by the recirculation created in front of the windward building (wake interference flow regime). Further reducing spacing (H/W>0.7) results in the skimming flow regime, where a stable recirculation is developed inside the canyon and the ambient flow is decoupled from the street flow.

The degree of interference can be expressed as an "interference factor" defined as $F_D/F_{D\_isoltaed\_cube}$, i.e. the ratio between drag force of the target cube surrounded by cubes and the drag force generated by an isolated cube. There are thus three cases: 1) no interference (interference factor = 1); 2) Sheltering (interference factor < 1); 3) Amplification (interference factor > 1).

### *4.1.3 Measured drag distribution*

Fig. 7 shows the interference factor for the cases with a boundary layer generated with roughness elements in the fetch. The cases with a boundary layer generated with no roughness elements in the fetch exhibit a similar behaviour. Both the drag forces measured by the balance and the pressure difference are presented. The balance measured the total contribution to the drag due to the form drag and friction. By definition the drag based on the pressure difference is the form drag. The cube is a bluff body so we expect the form drag to dominate.

From the figure it can be argued that the standard load cell method and the pressure-derived method provided similar distribution of the drag force within the array. This provides confidence in the measurements. Only the target cube located at the first row showed a normalized drag force based on the pressure about 10% larger than the drag force based on measurements by the balance. It is likely that this difference is due to that the pressure is measured in a finite number of points. Secondly, results confirm that the change in distribution of the drag force when changing the building packing density is in accordance with assessment of the distribution of the drag force based on measurements of the total drag force of the whole array reported in (Buccolieri et al., 2017). For further details see also the next subsection.

At the lowest packing density $\lambda_p$=0.028 (air gap between the cubes is 5$H$) there is no interference. And at $\lambda_p$=0.11 (air gap 2$H$) the interference factor is 0.5, while at $\lambda_p$=0.25 (air gap 1$H$) the force is almost totally exerted by the first cube. As discussed in Buccolieri et al. (2017) the latter case corresponds to a maximum drag force generated by the whole array (see Fig. 6 of their paper). With further increase of $\lambda_p$, the drag force slightly decreases until it becomes almost constant. One can claim that at $\lambda_p$=0.25 the array start to behave as one single unit. The effect of an increase of the frontal area is in fact cancelled out by the reduction of mean wind velocity (see Buccolieri et al., 2010). At $\lambda_p$=0.44 the air gap is 0.5$H$ and the drag force on the cube located downstream of the cube at the front of the array becomes negative. We interpret this as that the second cube now is submerged (Gowda and Sitheeq, 1990) in the boundary layer generated by the cube at the front. However it remains to verify this by flow visualization. A further detail is that starting with $\lambda_p$=0.11 the drag force on the cube at the front is less than on an isolated cube. This we interpret as that the cube downstream affect the drag force exerted on the cube at the front. However for the largest packing density $\lambda_p$=0.69 (air gap 0.2$H$) there is an amplification of the drag force on the front cube in the row that we cannot explain. Associated with the change in the distribution of the drag force with changing packing densities there is a change in the air flow pattern. In Figure 7 of Buccolieri et al (2010) it is shown that starting from packing density $\lambda_p$=0.44 there are recirculation zones within the array.

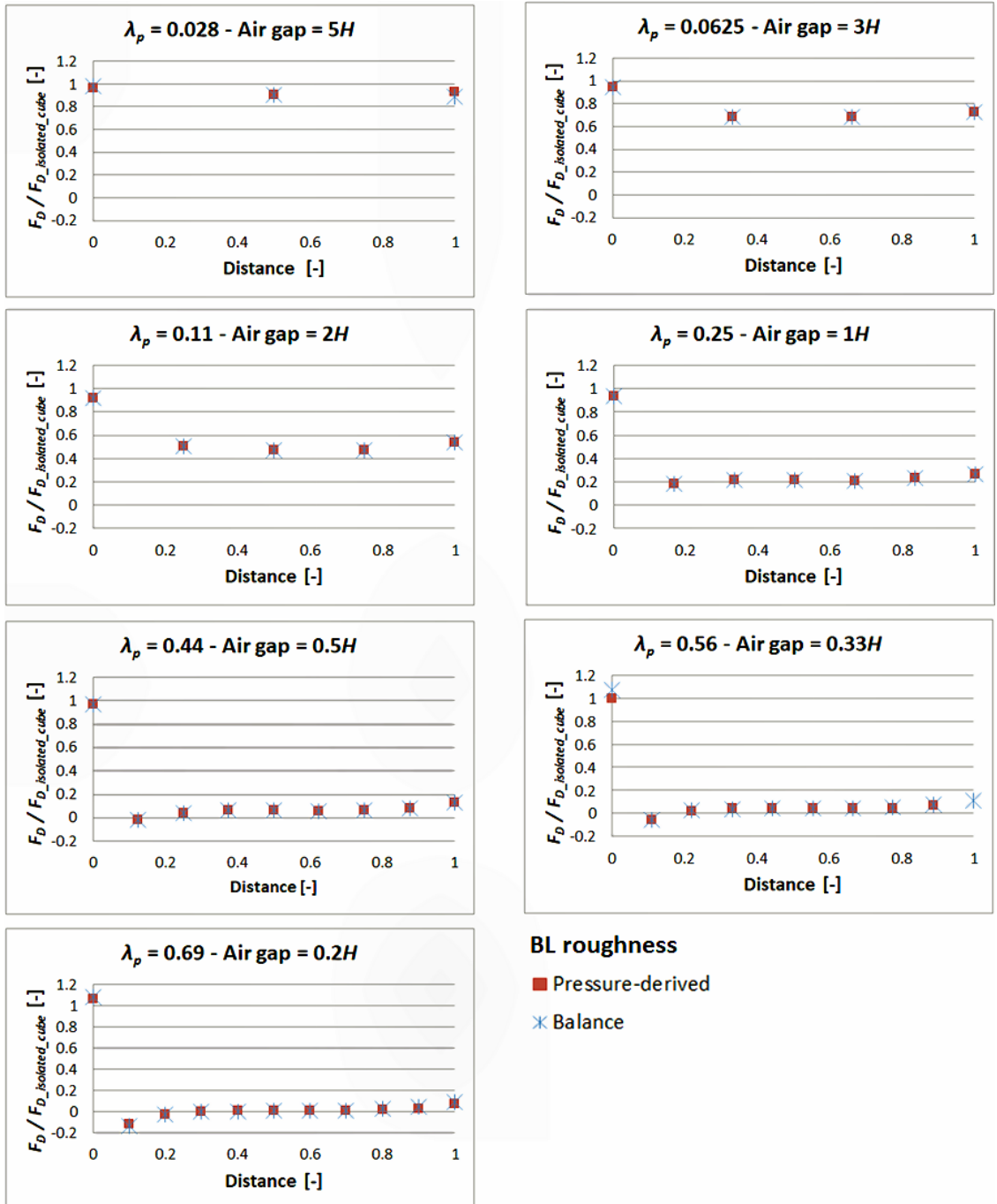

**Fig. 7.** Profiles of $F_D/F_{D\_isolated\_cube}$ (interference factor) generated by the target cubes, BL roughness case. The x-axis represents the distance from the first target cube of the array ("0") to the last target cube ("1"), see Fig. 1.

## 4.2. Assessment of the drag area

In Buccolieri et al (2017) the drag area distribution was assessed from the total drag force recorded over the whole array by calculating the ratio: Total drag area/Frontal area, i.e. the ratio between the total drag area retrieved from the total drag force and the physical frontal area of the whole array (see their Eq. 13).

Here the same relation is employed but using the drag force measured on the individual target cube. The total drag area is thus estimated by adding the drag forces measured with the balance generated by the single target cubes and the physical frontal area is the total frontal area of the target cubes. Two extreme cases of the frontal area are chosen, i.e., the physical frontal area of all the middle column of the array and the physical frontal area of the first (upstream) target cube. Each cube has a frontal area equal to $0.036 m^2$.

The comparison is shown in Fig. 8. First it can be noted that the drag distribution obtained from current measurements on individual cubes is similar to that obtained from measurements over the whole array. This suggests that, when estimating the drag area generating the drag force, the single middle column is representative of the whole array for each $\lambda_p$, indicating that our choice of isolated array well represents the drag force exerted by a portion of the city of a given $\lambda_p$. Second the figure shows that, as expected by the drag force distribution shown in Fig. 7, for low packing density the total frontal area is the most appropriate reference area (drag area) since the ratio is close to one, whereas for large packing densities the frontal area of the first row only is the one to be used as the appropriate reference area.

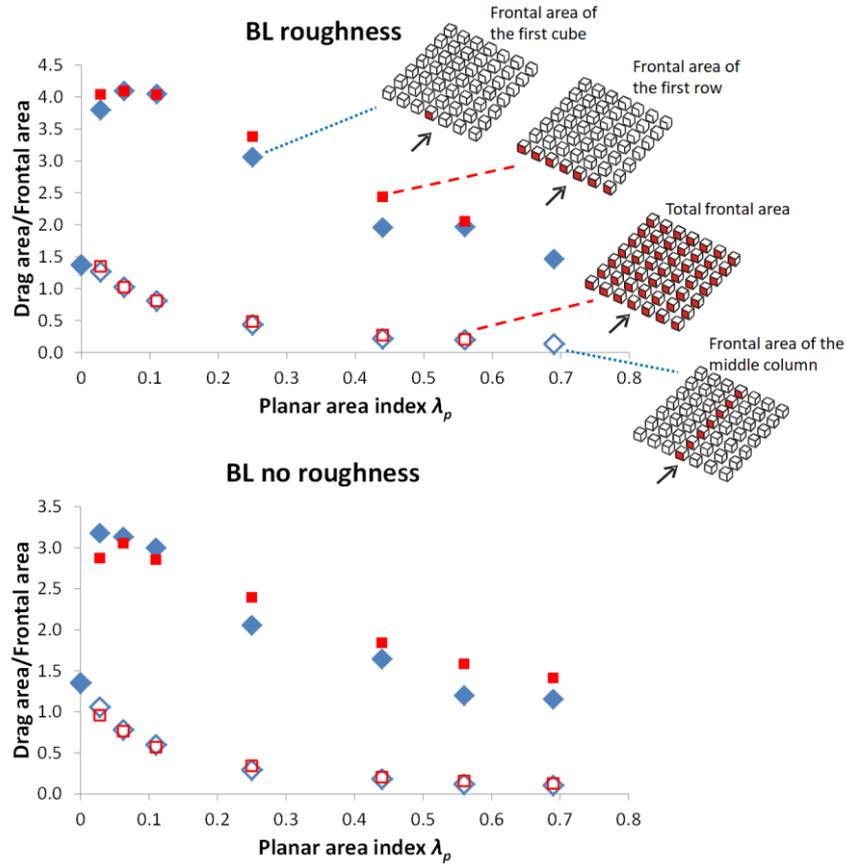

**Fig. 8.** Drag area/Frontal area as a function of the planar area index $\lambda_p$ when using the frontal area of the first cube or of the first row of cubes (full rhombus and square, respectively) and the frontal area of the middle column or of all the cubes (empty rhombus and square, respectively) in the a) ''BL roughness'' and b) ''BL no roughness'' cases. Note: square symbols refer to drag force measurements over the whole array (Buccolieri et al., 2017).

*4.3 Pressure distribution on the surface of target cubes*

To show the spatial distribution of pressure, Fig. 9 shows the pressure contours on the windward side of the first, central and last target cubes for the BL roughness case obtained by interpolating the measured pressure. Results for the BL no roughness follow a similar behaviour. We remind here that the pressure was measured (i) on the same target cube as the drag force was measured by the balance; and (ii) on one half of the side only (see Fig. 4), and therefore eventual asymmetries in the horizontal direction is not revealed which of course is a limitation. The stagnation point is the point with the highest pressure and from this point the air approaching the façade is distributed over the façade. For an isolated cube the stagnation point lies on the vertical symmetry line. However when a cube is lying in the front row of an array the location of the stagnation point is probably affected by the amount of air pressed into the street canyon, formed by two neighboring rows.

Fig. 9 shows that for all packing densities the first cube (i.e. the cube at the upstream first row of the array) has qualitatively the same pressure distribution as for the isolated cube. Further, the pressure distribution for the lowest packing density ($\lambda_p$=0.028) is qualitatively similar for all cubes in the row, that is the interference between the cubes is low in this case. This is in contrast to the case with the highest packing density ($\lambda_p$=0.69) where the pressure distribution on the cube at the centre and at the end of the row is almost uniform. The general trend is that with increasing packing density the gradient of the pressure distribution on the wall diminishes for the central and last cubes. This is due to the fact that with increasing packing density the width of the air gap between the cubes diminishes and this constrains the air motion. Because the pressure distribution is a footprint of the air motion along the wall, this results in a more uniform pressure distribution.

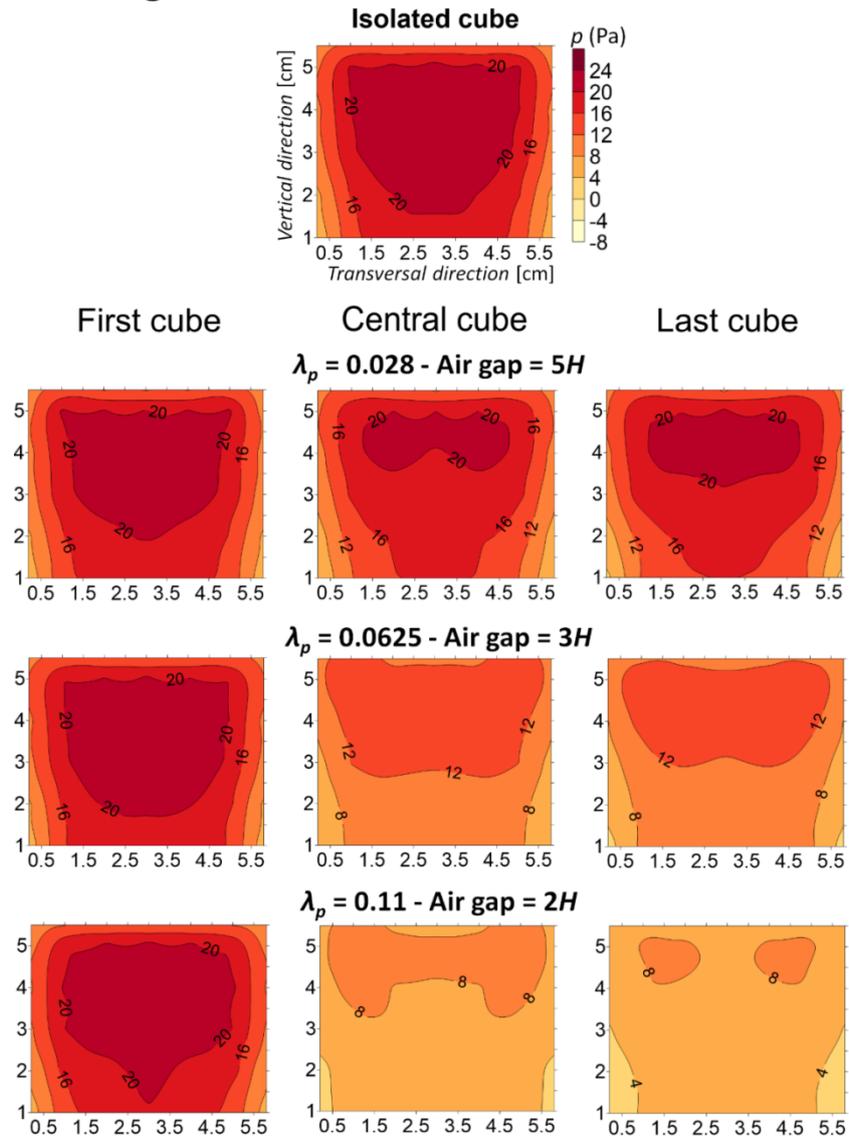

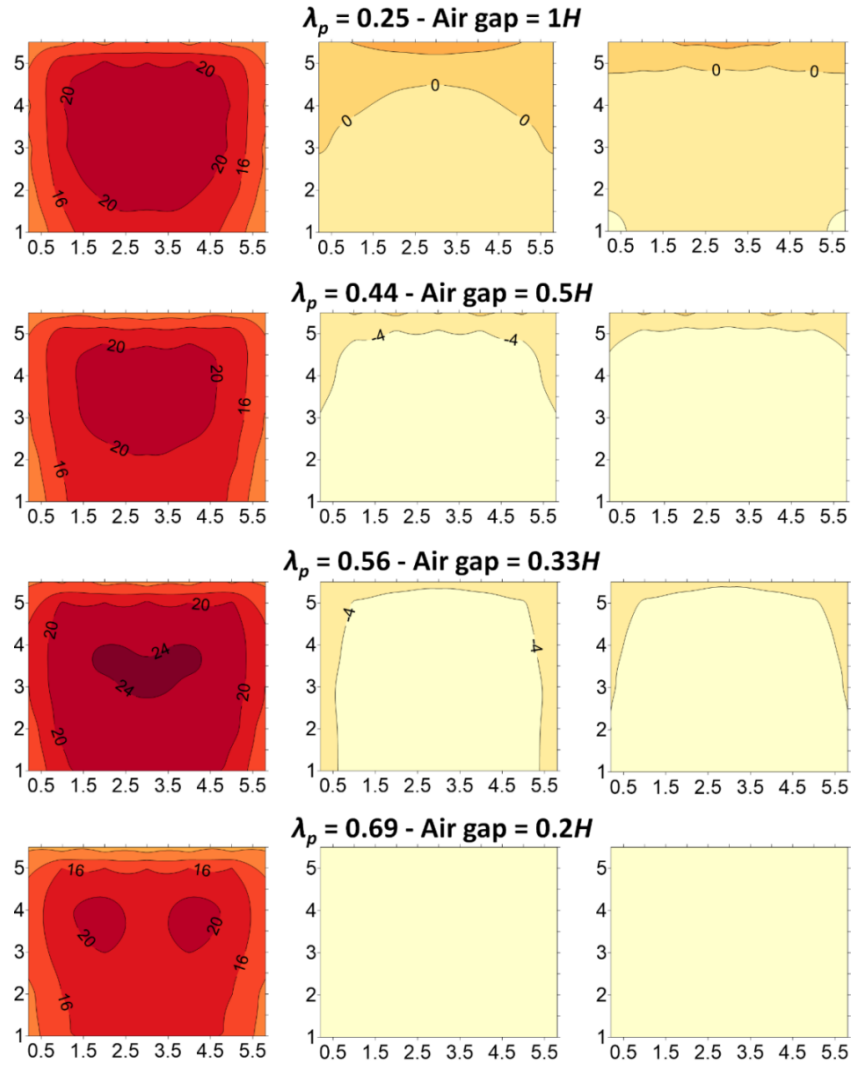

**Fig. 9.** Pressure contours on the windward façade of the first, central and last target cubes for each packing density $\lambda_p$, BL roughness case.

### *4.4. Assessment of the potential for cross ventilation*

The potential for cross ventilation is shown in Fig. 10 as the ratio between the velocity at the opening $U_{opening}$ and the reference velocity $U(H)$ at the height of the cube according to Eq. 20. This velocity ratio is dependent on the square root of the drag force multiplied with an expression which only varies with the type of boundary layer, generated either with no roughness elements in the fetch or with roughness elements in the fetch. We remind here that in Eq. 20 the force measured by the balance $F_D$(balance) is employed.

The figure shows that for the first (upstream) target cube the velocity ratio becomes about the same for both types of boundary layers and is approximately about 0.8.

Overall for cubes located downstream the velocity ratio with increasing packing density shows a similar qualitative behaviour as that found for the drag force (see Fig. 7). Specifically, when the packing density is $λ_p$=0.25 the velocity ratio has dropped to approximately 40% and by further increasing the packing density to $λ_p$=0.44 the velocity ratio drops to 20%. This shows that, as expected, cross ventilation is very much affected by the packing density. One can expect that single sided ventilation is less sensitive to the packing density because the ventilation is mainly generated by fluctuations. On the other hand single sided ventilation is in general much less than cross ventilation. For the same object the relation between cross ventilation and single sided ventilation has been studied in wind tunnel tests reported in Hayati et al. (2018).

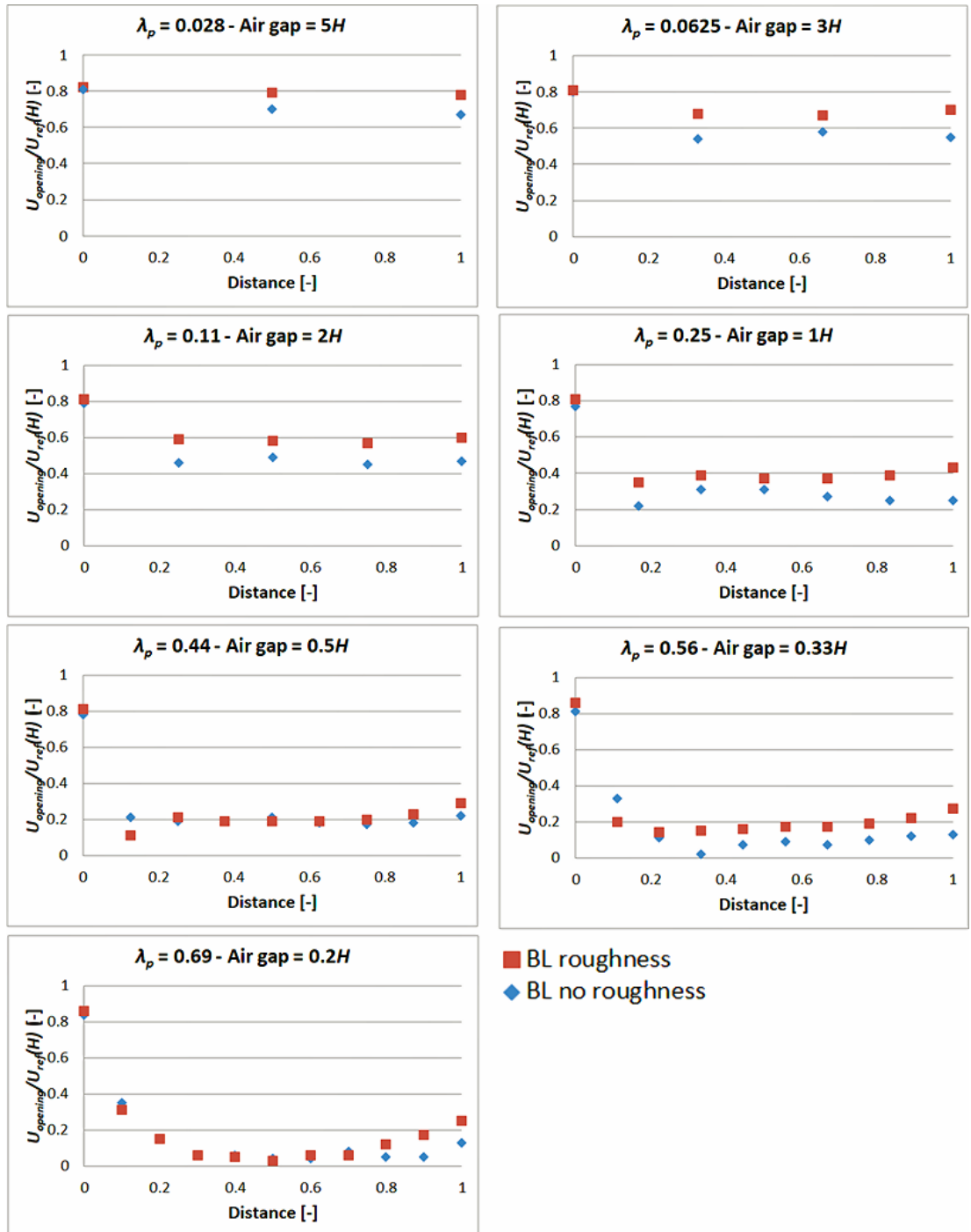

**Fig. 10.** Cross ventilation assessment through the ratio of velocity at the opening $U_{opening}$ and the reference velocity $U(H)$. The x-axis represents the distance from the first target cube of the array ("0") to the last target cube ("1") , see Fig. 1.

## 5. Conclusions

The drag force distribution generated by regular arrays of cubes was measured in a wind tunnel. The force was measured using both a standard load cell and indirectly estimated by measuring the static pressure at windward and leeward façades. The measurements with the two methods coincide within 10%. The drag force is presented as an interference factor by dividing the measured drag force with the drag force from an isolated cube. The main findings of the paper are summarized below:

- for the lowest packing density, the drag force is almost uniformly (within 15%) distributed between the cubes within the array. With increasing packing density the force exerted on the fraction of the cubes located downstream of the first row progressively diminishes, and for the highest packing densities the whole force is exerted on the first row, primarily. At the largest packing densities the force on the second row is negative i.e. directed opposite to the wind direction;

- as the packing density increases the interaction between the buildings changes from a collection of weakly interfering rows to become a single array, and finally as one row only. This implies that for very low packing densities the total frontal area is the most appropriate reference area (drag area) whereas for large packing densities the most appropriate reference area is the frontal area of the first row;

- the methodology previously presented in Buccolieri et al (2017) is further verified demonstrating that the distribution of the drag force within the array can be derived from the total drag force, using a combination of the total drag force expressed as a drag area and the physical frontal area of the cubes.

- the potential for cross ventilation is quantified as the velocity through a ventilation opening which is proportional to the square root of drag force. It is shown that there is a direct relation between the velocity through a ventilation opening and the in canopy velocity. This is an observation that has not previously pointed out.

 - the recorded pressure contours on the façades of the cubes is a footprint of the air motions in the air gap between the cubes showing that with increasing packing densities the gradient of the pressure decreases.

   Measuring the drag force correctly is not only relevant for the field of wind load on structures, but also for the derivation of improved description of the effect of the city within atmospheric mesoscale models. Several mesoscale studies of the drag force generated by arrays of buildings have been carried out with different methods. For example, Gutierrez et al. (2015) implemented a mechanical drag coefficient formulation depending on packing density following Santiago et al. (2010) into the Building Effect Parameterization + Building Energy Model system coupled with mesoscale Weather

Research Forecasting model. The mesoscale model was applied over New York obtaining an improvement of accuracy of mesoscale model in predicting surface wind speed in complex urban area. We expect that in the future the drag force distribution obtained for different wind directions (providing a sort of "drag force rose"), which may lead to very significant changes in the total drag imposed by the surface (Claus et al., 2012a,b; Santiago et al., 2013), can be the basis for a first order modelling of the dispersion of pollutants within an urban area.

**Acknowledgements**

Authors would like to thank Mr. Leif Claesson from the University of Gävle for his help in carrying out the wind tunnel experiments. One of the authors (SDS) kindly acknowledges the iSCAPE (Improving Smart Control of Air Pollution in Europe) project, which is funded by the European Community's H2020 Programme (H2020-SC5-04-2015) under the Grant Agreement No. 689954.